\documentclass[]{pasj01}

\begin{document} 
\Received{}
\Accepted{}

\title{X-ray upper limits of GW150914 with MAXI}


\author{Nobuyuki~\textsc{Kawai}\altaffilmark{1,2}}
\altaffiltext{1}
{Department of Physics, Tokyo Institute of Technology, 2-12-1 Ookayama, Meguro-ku, Tokyo 152-8551}
\altaffiltext{2}
{MAXI team, RIKEN, 2-1 Hirosawa, Wako, Saitama 351-0198, Japan}

\author{Hitoshi~\textsc{Negoro}\altaffilmark{3}}
\altaffiltext{3}{Department of Physics, Nihon University, 1-8-14 Kanda-Surugadai, Chiyoda-ku, Tokyo 101-8308, Japan}

\author{Motoko~\textsc{Serino}\altaffilmark{2}}%

\author{Tatehiro~\textsc{Mihara}\altaffilmark{2}}%

\author{Kazuki~\textsc{Tanaka}\altaffilmark{3}}%

\author{Takahiro~\textsc{Masumitsu}\altaffilmark{3}}%

\author{Satoshi~\textsc{Nakahira}\altaffilmark{2}}%


\maketitle

\begin{abstract}
We searched for X-ray candidates of the gravitational wave (GW) event
GW150914 with Monitor of All-sky X-ray Image (MAXI).
MAXI observed the error region of the GW event
GW150914 from 4 minutes after the event and covered about 90\%
of the error region in 25 minutes.
No significant time variations on timescales of 1 s to 4 days were found
in the GW error region.
The $3\sigma$ upper limits for the X-ray emission associated with the GW event
in 2--20 keV were 
9.5 $\times 10^{-10}$, 2.3 $\times 10^{-10}$, and 0.8 $\times 10^{-10}$ 
ergs cm$^{-2}$ s$^{-1}$ 
for the time scale of $\sim$ 1000 s, 1 day, and 10 days,
respectively.
If GW events are associated with short GRBs like GRB 050709, 
MAXI will be able to detect X-ray emissions from the source.
\end{abstract}

\section{Introduction}


The first detection of the gravitational wave (GW) has been made
by LIGO on September 14, 2015, as known as GW150914
\citep{2016PhRvL.116f1102A}.
The strain waveform indicates merger of two black holes (BHs) with
masses of 29 and 36 $M_{\odot}$. It is the first test of the general 
relativity in the strong field. It is also the first test of the theory
for propagation of the gravitational wave in the space.
A rough distance of $\sim$400 Mpc was derived by the waveform analysis
with a rather large uncertainty. Another important aspect
of this event is the first solid evidence for existence of BHs
with intermediate masses $\sim$ 30$M_{\odot}$ , and the resulting BH with 
60$M_{\odot}$ .
Reliable dynamical mass measurements of BHs have
been made only for the stellar mass BHs in the 
Milky Way galaxy and the Large Magellanic Cloud,
 and the super-massive BHs that lie at centres of
galaxies including the Milky Way. 
The stellar mass BHs whose
masses are smaller than a few tens of $M_{\odot}$ are known to be produced
by collapse of massive stars at the end of their lives. However, origin
of supermassive BHs in nuclei of galaxies is not yet clear.
They may be produced by hierarchical mergers of many BHs, 
or by accretion of gas to the single BH. In 
either case the massive BHs at high redshifts indicates that
there need to be more massive BH as the seed. 
\citet{2014MNRAS.442.2963K} 
suggested that BHs of $\sim$ 30$M_{\odot}$ are 
naturally produced by collapse of pop-III stars at their endpoints, 
and that binaries of such BHs are the most common sources of
gravitational waves. GW150914 is naturally explained in this
scenario, and may imply that the remnants of the Universe's first
stars may be found in the neighbourhood of the Milky Way. In
order to confirm or test these theories on the birth and evolution
of massive BHs, the distance and the environment
of the BHs mergers are essential. The poor localization
of GW does not allow us to associate the source to any known
class of objects such as galaxies, or to know its relative position to its
host galaxy.

In order for that we need precise location that is only achievable
with electromagnetic (EM) waves. EM counterparts
of gravitational wave event have been discussed quite extensively
for the case of merger of double neutron star (NS) binary,
or NS-BH binaries, where the ejected NS material are supposed to produce 
EM emission through nuclear decay of r-process elements (``kilo-nova'';
\cite{2010MNRAS.406.2650M}),
accretion on to the newly formed BH
\citep{2011Natur.478...82N,2013MNRAS.430.2121P}, 
or interaction of Blandford-Znajek jet from the rotating BH
\citep{2014ApJ...796...13N}.

Not so much discussion was made for BH pairs,
but some mechanism have been suggested for possible production 
of EM emission. For example, 
\citet{2016PhRvD..93d4048N} 
suggests 
a mechanism in which the merged BH could accrete from 
the interstellar medium to emerge as an EM counterpart.

The possible gamma-ray detection \citep{2016ApJ...826L...6C} 
resembling a weak short
gamma-ray burst have prompted theoretical ideas for EM emission.
It may not be at all impossible for BH mergers to
produce EM emission, if the environment is suitable.

In this paper we present the MAXI follow-up of GW150914.
MAXI (Monitor of All-sky X-ray Image; \cite{2009PASJ...61..999M})
is an X-ray all-sky monitor on the International Space
Station (ISS). 
It scans most of the sky in every orbit ($\sim$ 92min) of the
ISS with its narrow and long field of view. Most of the error
region of GW150914 was covered by MAXI following the event, 
and placed upper limits on the X-ray emission from the GW event 
on various time scales.

\section{Instrumentation}
MAXI has two instruments: GSC (2--20 keV; \cite{2011PASJ...63S.623M}) 
and SSC (1--7 keV; \cite{2011PASJ...63..397T}).
The instant field of views (FOVs) of GSC and SSC are about 2\% and 1\% 
of the whole sky.
The FOVs scan the whole sky once in 92 minutes.
Currently 6 out of 12 GSC cameras are functioning \citep{2014SPIE.9144E..1OM}.
GSC are not operating in the regions with high particle-background,
which are South Atlantic  Anomaly and higher latitude than $\sim 40$ degrees.
The functioning time is about 40\%.
GSC is turned off in the vicinity of the sun ($\sim 5$ degrees).
Still, GSC can cover about 85\% of the whole sky in 92 minutes
\citep{2011PASJ...63S.635S}.
Because SSC is operated in the night time to avoid the sun light,
its operating efficiency becomes considerably low.
The SSC functioning time and sky coverage in 92 minutes
are about 25-30\% and 30\%, respectively.

MAXI/GSC is capable to detect transient events 
 with the  limit of $\sim$2 $\times 10^{-9}$ erg cm$^{-2}$ s$^{-1}$ in the 2--20 keV band
(e.g. \cite{2014PASJ...66...87S,2016PASJ...68S...1N}) in a scan transit.

\section{Observations}
 \subsection{time and area of the observation}
\begin{figure}
 \begin{center}
   \includegraphics[width=8cm]{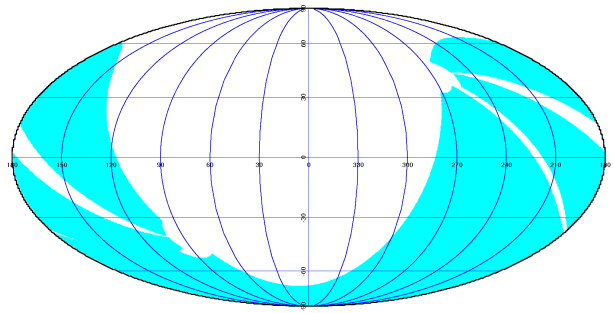}
 \end{center}
\caption{The schematic picture of the observed region from 
  $t0$ + 4 min to $t0$ + 25 min ({\it the first scan})
  by GSC (cyan).
  The figure is shown in the equatorial coordinates.
      }
\label{fig:allsky_mol}
\end{figure}

\begin{figure}
 \begin{center}
  \includegraphics[width=8cm]{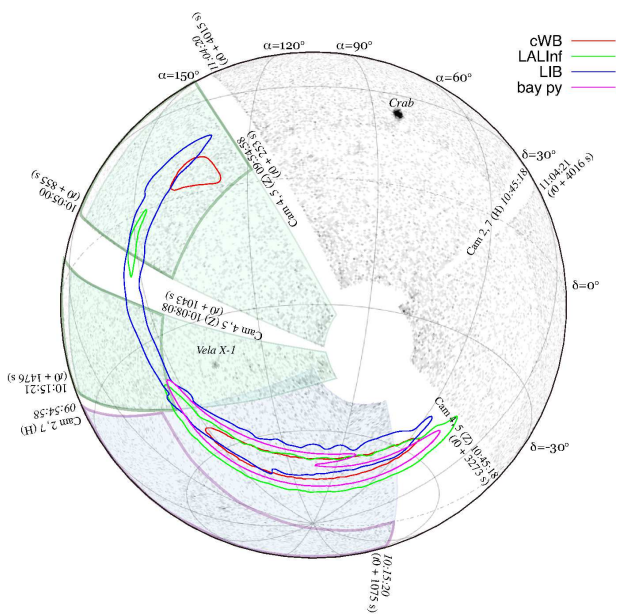}
 \end{center}
\caption{The schematic picture of the observed area with GSC
  from $t0$ + 4 min to $t0$ + 74 min
  with GW 90\% probability contours 
   by various algorithms 
   	(cWB, LALInf, LIB, and bay py). 
	Regions observed with GSC\_4 and GSC\_7 are 
	surrounded by a green bold-line and a purple one, respectively. 
        Most of the visible parts
	on this image of the regions are also observed with 
        GSC\_5 and GSC\_2, respectively. 
      }
\label{fig:obs_hankyu}
\end{figure}

  MAXI observes a point of the sky every $\sim$92 minutes.
  The first GSC observation of the GW150914 region carried out from
  $t0$ (=2015/09/14 09:50:45 UTC) + 4 min to $t0$ + 25 min 
  ({\it the first scan}, hereafter).
  Figure \ref{fig:allsky_mol} shows the sky map of the observed area by
  GSC during {\it the first scan}.
  Figure \ref{fig:obs_hankyu} shows the observed area and scan time
  of GSC from $t0$ + 4 min to $t0$ + 74 min ({\it the first orbit}).
  Probability maps of the GW source position were calculated
  by various algorithms: 
   Coherent Wave Burst (cWB; \cite{2016PhRvD..93d2004K}), 
   LALInference (LALInf; \cite{2015PhRvD..91d2003V}), 
   LALInference Burst (LIB; \cite{2015arXiv151105955L}), 
   and 
   BAYESTAR (bay py; \cite{2016PhRvD..93b4013S}). 
  The region with high significance are observed mainly by GSC\_2,
  GSC\_4, and GSC\_5.

\begin{figure*}
 \begin{center}
 \includegraphics[width=12cm]{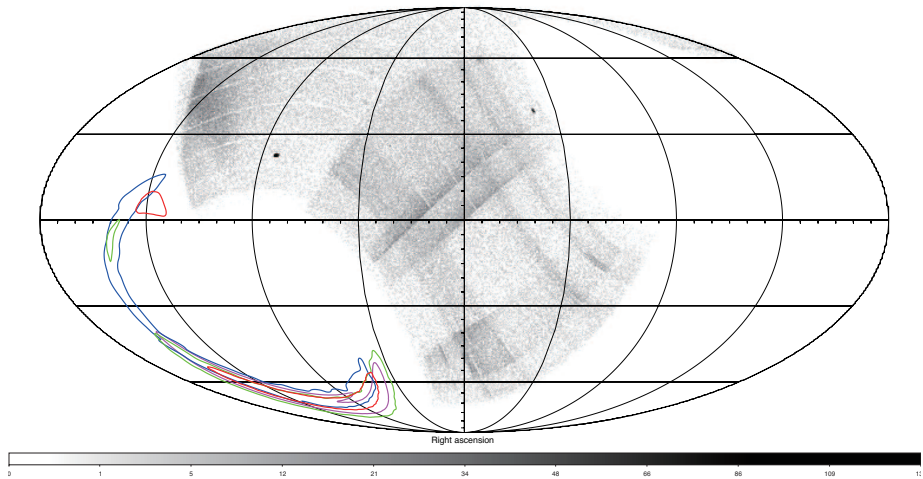}
 \end{center}
\caption{A single pixel event X-ray image observed by SSC from 
  $t0$ + 48 min to $t0$+1day.
The GW 90\% probability contours with the same colors in figure \ref{fig:obs_hankyu} were also shown.
      }
\label{fig:ssc1day}
\end{figure*}

  The SSC observation did not started until $t0$+48 min,
  since the ISS entered the day-earth region from $t0$-12 min to $t0$+44 min, 
  and also South Atlantic Anomaly from $t0$+35 min to $t0$+46 min,
  Figure \ref{fig:ssc1day} shows the all-sky image obtained by 
  two SSC cameras in 1 day.
  
\begin{figure*}
 \begin{center}
  \includegraphics[width=12cm]{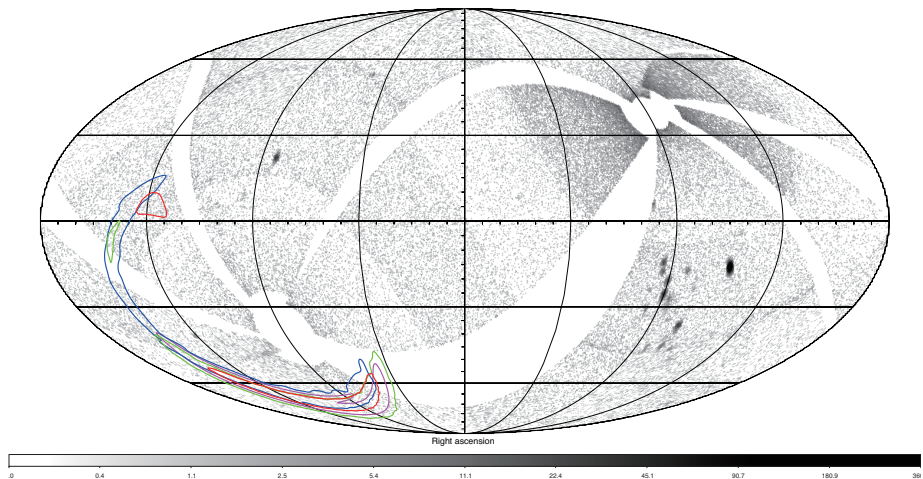}
 \end{center}
\caption{An X-ray image observed by GSC 
  from $t0$ + 4 min to $t0$ + 74 min.
 GW contours are same as figure \ref{fig:ssc1day}.
      }
\label{fig:gsc1scan}
\end{figure*}

\begin{figure*}
 \begin{center}
  \includegraphics[width=12cm]{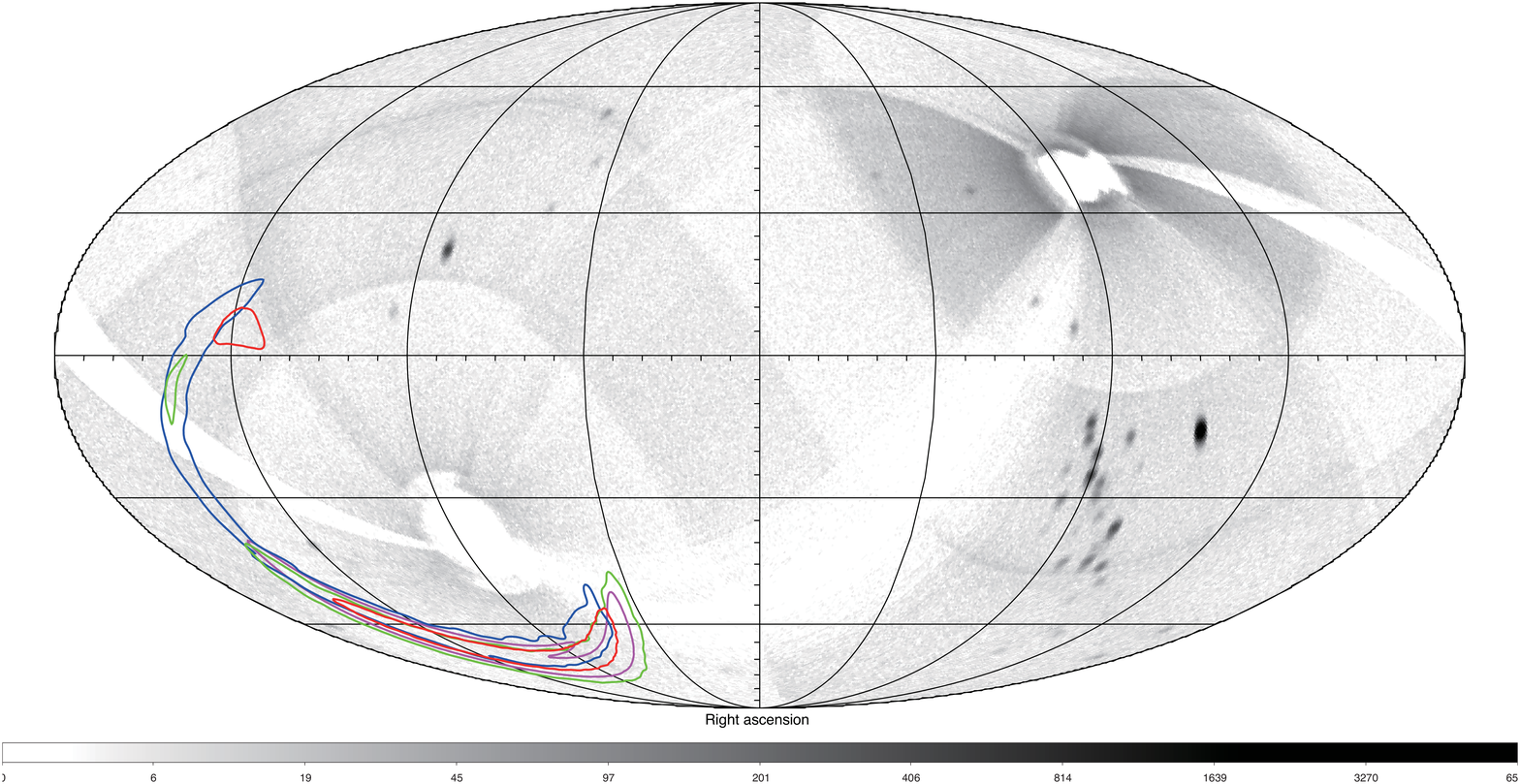}
 \end{center}
\caption{An X-ray image observed by GSC from 
  $t0$ + 4 min to $t0$+1day.
 GW contours are same as figure \ref{fig:ssc1day}.
      }
\label{fig:gsc1day}
\end{figure*}

 \subsection{coverage}
  We calculated the coverage of the 90 percentile region of 
  each GW skymap by the following procedure.
  First, we calculated the HEALPix map%
  \footnote{the pixel size of this GSC map is $\sim$ 1 deg $\times$ 1 deg}
  of the region 
  which is observed by each GSC camera during {\it the first scan}
  (figure \ref{fig:allsky_mol}).
  If the center of a pixel is in the field of view of the camera
  during the time, we regard the pixel as an observed one.
  Since the pixel size is smaller than the GSC point spread function,
  it is reasonable.
  Then we add the maps of all cameras.

  Next, we listed the pixel number in the 90 percentile region
  of each HEALPix map of GW. Then examined in the MAXI map 
  whether the pixels were observed or not.
  As a result, we obtained the observation coverage within {\it the first scan}
  for each GW map (table \ref{tab:cover}).

 We also show all-sky X-ray images obtained by working 6 GSC cameras
  from $t0$ + 4 min to $t0$ + 74 min 
  (in 1.5 hours; figure \ref{fig:gsc1scan}),
 and in 4 days (figure  \ref{fig:gsc1day}).
In producing these GSC images,  we did not use GSC\_3
data of the region where the GSC\_4 had also observed. 
This was because the background rate in GSC\_3 was high 
due to the loss of the anti-coincident background rejection function
\citep{2014SPIE.9144E..1OM}. 

We also did not use the events detected with the anode 5 of 
the GSC\_5
 because solar soft X-rays made a fake point-like image around 
 the region $(\alpha, \delta) =(159, 16)$.
As can be seen from these images, GSC observed about 80\% and 95\% of 
the whole sky in 1.5 hours
and in 1 day, respectively.
SSC covered 39\% of the whole sky in 1 day (figure \ref{fig:ssc1day}).

\begin{table}
  \tbl{observation coverage of the GW maps
  in {\it the first scan} ($t0$ + 4 min to $t0$ + 25 min).}{%
  \begin{tabular}{llll}
    \hline
     map   & pix. 90\%\footnotemark[$*$] 
      &  pix.obs.\footnotemark[$\dagger$] 
       & coverage\footnotemark[$\ddagger$] \\
    \hline
     LALInf  & 45863 & 38600  &  84.2\% \\
     LIB     & 56898 & 50451  &  88.7\% \\
     cWB     &  1469 &  1395  &  95.0\% \\
     bay py  & 33440 & 30702  &  91.8\% \\
    \hline
  \end{tabular}}\label{tab:cover}
  \begin{tabnote}
    \footnotemark[$*$] 
      number of the HEALPix pixels contained in the 90 percentile
      region of each map \\
    \footnotemark[$\dagger$]
      number of the pixels contained in the 90 percentile
      region of each map and observed by MAXI \\
    \footnotemark[$\ddagger$]
      observation coverage of 90 percentile region by MAXI
  \end{tabnote}
\end{table}

 \subsection{Event search by the nova-alert system}
 
  The MAXI nova-alert system \citep{2016PASJ...68S...1N} detected no significant
  time variability in any error regions for 4 days since the GW trigger time.
The nova-alert system consists of a nova-search system(s) to find time variability 
and an alert system to eliminate further statistical significance of the events triggering the nova-search system.

\begin{figure}
 \begin{center}
  \includegraphics[width=8cm]{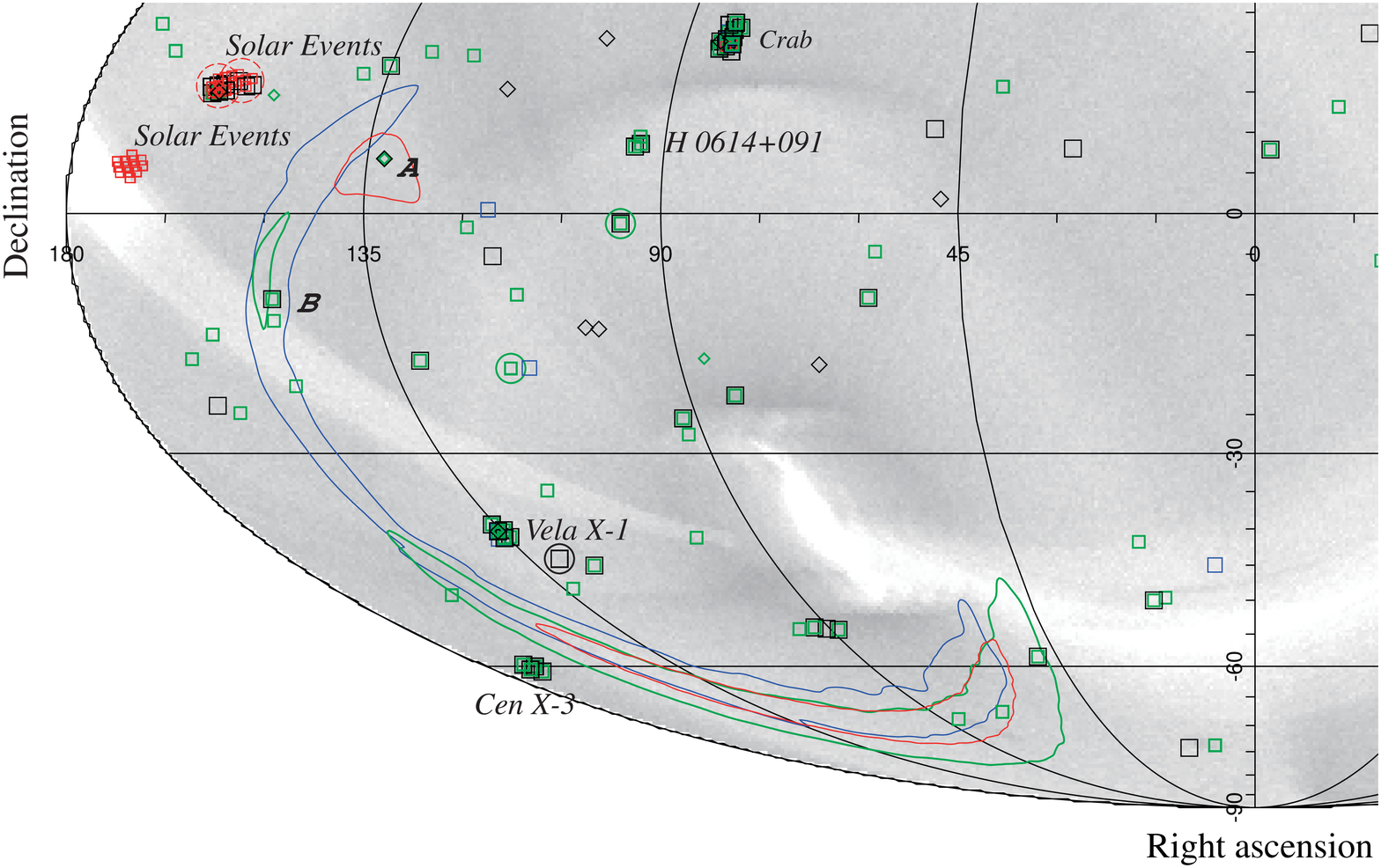}
 \end{center}
\caption{Locations of the triggered events in a nova-search system in the first orbit 
(shown by triangles) and in 4 days (squares), and those of 
more significantly detected events (circles, see text for more detail), 
overwriting on the 4-day GSC 2-20 keV image. 
Only the 90\% probability contours are shown, and the bay-py contour is 
omitted to avoid complexity. 
Events shown in red (and black) in the upper left side are due to solar X-rays.}
\label{fig:nsdetection}
\end{figure}

In figure \ref{fig:nsdetection}, we plot locations of triggered events in one of two nova-search systems
(a system with relatively high event thresholds) 
for the first 4 days around the error regions.
 The diamonds represent short-term events that triggered in 1.5 hours
 in 1 s, 3 s, 10 s, 30 s, and 1 orbit ($\simeq 92$ min) integrated-time bins.
The squares show long-term events
triggered from $t0 $ + 1 orbit to $t0$ + 4 orbits (in 4 orbits bin), from $t0$ + 4 orbits to $t0$ + 1 day (1 d bin)
and from $t0$ + 1 day to $t0$ + 4 days (4 d bin).
 The colors, black, red, green, and blue, of the marks  represent
 energy bands triggered, corresponding to the 3--10 keV, 2--4 keV, 4--10 keV, and 10--20 keV
energy band, respectively. 
Different mark sizes for different energy band data are to 
avoid overwriting, and does not show any significance.
 Chance probabilities to trigger, i.e., the trigger criteria, are 
 $\le 10^{-3}$ to $10^{-4}$. 

The circles show {\it detected} events, related with the triggered events for 4 days, 
which meet {\it detected} criteria as statistically 
significant events in the alert system \citep{2016PASJ...68S...1N}.
As described previously, no event was detected in any of the
90\% probability regions.
Bright catalogued sources, such as Cen X-3 and Vela X-1, and their neighborhoods often triggered the system, 
but are masked in the alert system.

Triggered events without circles are not statistically significant 
(usually at less than $3\sigma$ levels), but are candidates of variable events.
An event {\bf A} at $(\alpha, \delta) =(132.43, 6.73)$ 
in the cWB error regions is noticeable because it first triggered 
at 09:57:42 ($t0 + 417$ s) in 30-s and 1-orbit time bins. 
The 4--10 keV flux at the region in the scan transit at 09:57 
was $0.035^{+0.018}_{-0.015}$ counts cm$^{-2}$ s$^{-1}$, and
we could not confirm any point-source like excess for this event in GSC images.

An event B at $(\alpha, \delta) = (150.44, -10.55)$ near the 
LALInf 90\% probability region and in the LIB one triggered the system
from 11:37:36 ($\sim t0 + 107$ min) from 13:10:19 ($\sim t0+200$ min) in the 4 orbits bin.
These events, however, are due to the noise caused by the reduction of 
high voltage of the counters 
and not astronomical events.

 \subsection{upper limits of the flux}
  We evaluated the upper limits of the flux by the following procedure.
  First, we selected 10 points representing the observed region and
  counted the photons in the circular regions with the radii of 
  1.5 deg, which is the typical size of the PSF.
  The 1-sigma fluctuation  of the background is defined as $\sqrt{n}$,
  where $n$ is observed count in the circular region.
  Next, we calculated the effective exposure $a$, which has the dimension
  of area $\times$ time, of each of the 10 points.
  Then we regarded $f \equiv 3 \, \sqrt{n} / a $ as 3-sigma upper limit
  of the flux at the point.
  The averages of 3-sigma upper limits of the points for the observations 
  of a scan is 0.12 $\pm$ 0.02 c s$^{-1}$ cm$^{-2}$ (in 2--20 keV), 
  which corresponds to the energy flux of 
  $(9.5 \pm 1.8) \times10^{-10}$ erg s$^{-1}$ cm$^{-2}$.
  The upper limits for the one day and ten days observations 
  in the same energy band are
  of a scan 0.029 $\pm$ 0.004 c s$^{-1}$ cm$^{-2}$
  [$(2.3 \pm 0.4) \times10^{-10}$ erg s$^{-1}$ cm$^{-2}$] and
  of a scan 0.010 $\pm$ 0.001 c s$^{-1}$ cm$^{-2}$
  [$(8.2 \pm 1.0) \times10^{-11}$ erg s$^{-1}$ cm$^{-2}$], respectively.

\label{ss:obs}

\section{discussion}

\begin{table*}
  \tbl{Upper limits for the X-ray flux and radiated energy obtained by MAXI/GSC.}{%
  \begin{tabular}{ccllll}
    \hline
     & Timescale & Flux (2--20 keV) & Luminosity \footnotemark[$*$] & Radiated & $E_X/E_{GW}$ \\
     & (s) & (erg s$^{-1}$ cm$^{-2}$) &  (erg s$^{-1}$)  & Energy (erg) &  \\
    \hline
     1 orbit    & 1000 & $ <9.5\times 10^{-10}$ & $<1.9\times 10^{46} $ & $ <1.9\times 10^{49} $ & $ <3.5\times 10^{-6} $ \\
     1 day    & $ 8.6\times 10^4$ & $ <2.3\times 10^{-10} $ & $ <4.6\times 10^{45} $ & $ <4.0\times 10^{50} $ & $ <7.4\times 10^{-5}$  \\
   10 days   & $ 8.6\times 10^5$ & $ <0.8\times 10^{-10} $ & $ <1.6\times 10^{45} $ & $ <1.4\times 10^{51} $ & $ <2.6\times 10^{-4}$  \\
    \hline
  \end{tabular}}\label{tab:energy}
  \begin{tabnote}
    \footnotemark[$*$] 
	Distance of 410 Mpc assumed
  \end{tabnote}
\end{table*}

\begin{figure}
 \begin{center}
   \includegraphics[width=8cm]{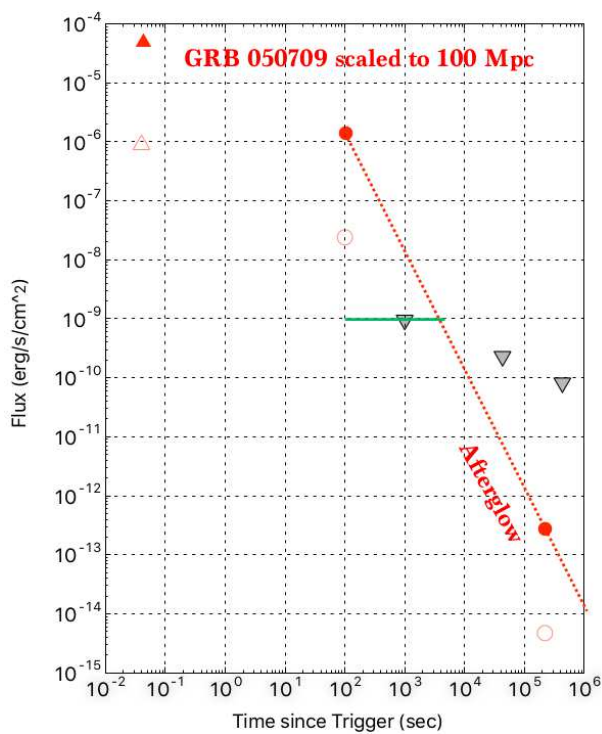}
 \end{center}
\caption{The upper limits for the 2--20 keV X-ray flux associated with GW150914 on three time scales, 1000s, 1 day, and 10 days, are shown with gray down-pointing triangles.  
They are compared with the X-ray fluxes of GRB050709 at three different phases: short pulse, extended emission, and afterglow shown by open triangle and circles.
These fluxes are then scaled to the source distance of 100 Mpc and plotted with filled symbols.
The dashed line indicate a possible afterglow light curve connecting the soft X-ray extended emission and the late afterglow.
      }
\label{fig:ul+grb050709}
\end{figure}

The upper limits  on the X-ray flux on three different time scales can be summarized in table \ref{tab:energy}.
The upper limits for the energy radiated in X-ray over the measurement time scale are also shown, 
and compared with the energy radiated in gravitational wave.
These can be treated as upper limits for the extended X-ray afterglow of the GW event.
For example, we estimate that the energy radiated in X-ray afterglow over 1000 seconds 
is less than $3.5 \times 10^{-6}$ of the total energy released in the BH merger.
Since GW150914 is a BH binary merger,  strong X-ray emission is not naturally expected.
In that respect, our upper limits do not constrain the theory or information on the environment.
On the other hand, possible detection of a weak short gamma-ray transient similar to a short GRB by Fermi GBM was reported
\citep{2016ApJ...826L...6C}. 
The future observation of LIGO with improved sensitivity
is expected to detect gravitational waves from mergers of double NSs or
BH-NS binaries are expected.
For these systems, significant fraction of NS matter is expected to be ejected.
Theories predicts various ways of generating electromagnetic radiation from these events
resulting from the radiative decays of r-process nuclei (kilonova) and free neutrons,
or  the relativistic jet powered by the central engine such as accreting BH or rotating magnetic compact objects through the Blandford-Znajek process.
In particular, the latter is considered as the promising origin of short gamma-ray bursts, that emits intense gamma-ray radiation followed by extended X-ray emission.

It is interesting to compare the MAXI sensitivity to the X-ray flux expected from the possible gamma-ray transient recorded by Fermi GBM.
Using the photon power-law index $-1.4$ and fluence between 10 and 1000 keV of $2.4 \times 10^{-7} {\rm erg\; cm}^{-2}$  reported for this event, 
we estimate the 2--20 keV fluence of  $\sim 2 \times 10^{-8}$ erg cm$^{-2}$.
If the event were captured at the middle of the scan transit, 
it would produce a significant detection with $\sim$ 10 counts in a MAXI GSC camera in less than a second, 
which is more than an order of magnitude higher than the background of GSC \citep{2011PASJ...63S.635S} for that duration.
In general, however, MAXI has only a small chance for detecting prompt emission coincident with the gravitational wave, 
because the instantaneous sky coverage of MAXI is only 2\% of the entire sky.
We need to wait for more GW events to be detected for such a luck.
 
It is also instructive to see how the present MAXI upper limits for GW150914
are compared with a possible future detection of a short GRB coincident with GW detection.
In Fig. \ref{fig:ul+grb050709} we plot the MAXI upper limits for the X-ray flux as a function of the time since the GW150914 trigger.
The three points are naturally aligned on a straight line on a logarithmic plot following a $\propto t^{-1/2}$ relation 
expected for the background-limited sensitivity.

We can also compare the MAXI sensitivity with the X-ray flux of GRB 050709, a short GRB.
We choose this short GRB for comparison, since it is the only short GRB for which the prompt burst phase has been observed in the energy band common with MAXI/GSC.
No other GRB missions like Swift and Fermi have comparable sensitivity in the X-ray band below 10 keV.
The WXM on HETE-2 observed the prompt short-hard pulse with duration $\sim$0.3 s and the extended soft X-ray emission that lasted $>$100 s
in the 2--25 keV X-ray band \citep{2005Natur.437..855V}.
Its X-ray afterglow was detected by Chandra, which lead to Hubble detection of optical afterglow and identification of the host galaxy at $z\approx 0.16$ with the precise localization \citep{2005Natur.437..845F}.
We plot its X-ray fluxes in the short hard pulse, the extended X-ray emission, and the afterglow with open symbols in Fig. \ref{fig:ul+grb050709}.
These fluxes are scaled to the source distance of 100 Mpc, the expected range for double NS merger with LIGO O2 \citep{2016LRR....19....1A}.
It is immediately clear that the prompt X-ray emission, both short pulse and extended emission,  of GRB 050709 is far brighter than the detection threshold of MAXI/GSC even at its original redshift, not to mention the case scaled to the LIGO O2 range.
Here after we discuss the possibility for detecting the X-ray emission if a short GRB is associated with the NS merger event in the LIGO O2 run.
Despite this high flux, the chance for detecting short pulse is expected to be very low because of the narrow collimation of short GRBs emission (\cite{2015ApJ...815..102F}) and MAXI's small instantaneous sky coverage.
While there is strong evidence for the short pulse originating in a relativistic jet with small opening angle, the nature and origin of the soft extended emission remains a mystery.
If the collimation of the soft X-ray extended emission is weak, as in the model proposed by \citet{2014ApJ...796...13N}, the chance for MAXI detection may not be negligible. 
The MAXI sky coverage may be still a problem, since the duration of soft extended emission is much shorter than the scan interval of MAXI of 92 minutes, the ISS orbital period.
However, if the soft extended emission is connected to the late afterglow as indicated by a dashed line in Fig. \ref{fig:ul+grb050709}, 
its flux stays above the MAXI threshold for more than 3000 s, a major fraction of the scan interval, suggesting a higher probability for detection.

In summary, MAXI set an upper limit for the X-ray emission associated with the gravitational wave event GW150914 on the timescales of one orbit ( $\sim$ 1000 s), day, and 10 days following the GW trigger.
In the future GW observing runs, MAXI has possibility to constrain the model for electromagnetic radiation and association of GW events with short GRBs.


\begin{ack}
This research has made use of the MAXI data provided by RIKEN, JAXA 
and the MAXI team.
This research was supported by JSPS KAKENHI
Grant Number 24740186.
\end{ack}


\bibliographystyle{aa}
\bibliography{ref}

\end{document}